\documentclass[aps,twocolumn,showpacs,superscriptaddress]{revtex4}
\usepackage{epsfig}

\def\avg#1{\langle#1\rangle}
\def\Re{\rm{Re}}
\def\Im{\rm{Im}}
\def\be{\begin{equation}}       \def\ee{\end{equation}}
\def\bea{\begin{eqnarray}}      \def\eea{\end{eqnarray}}

\begin{document}

\title{
Thermodynamic properties of the $d$-density-wave order in cuprates
}

\author{Congjun Wu}
\affiliation{Department of Physics, University of Illinois at 
Urbarna-Champaign, 1110 West Green Street, Urbana, IL 61801}
\affiliation{Department of Physics, McCullough Building, 
Stanford University, Stanford CA 94305-4045}
\author{W. Vincent Liu}
\affiliation{Department of Physics, University of Illinois at 
Urbarna-Champaign, 1110 West Green Street, Urbana, IL 61801}
\affiliation{Center for Theoretical Physics, 
Department of Physics, Massachusetts Institute of
Technology, Cambridge, MA 02139}

\begin{abstract}
We solve a popular effective Hamiltonian of competing $d$-density wave
and $d$-wave superconductivity orders self-consistently at the mean-field
level for a wide range of doping and temperatures.  The theory predicts
a temperature dependence of the $d$-density wave order parameter
seemingly inconsistent with the neutron scattering and $\mu$SR
experiments of the cuprates.  We further calculate  
thermodynamic quantities, such as chemical potential, entropy and
specific heat.  Their distinct features  can be used to
test the existence of the $d$-density wave order in cuprates.
\end{abstract}
\maketitle

Unconventional charge and spin density wave orders were extensively 
investigated in correlated electron systems 
\cite{history}. 
Recently, Chakravarty {\it et al} \cite{char1}
proposed that the pseudogap phenomena in the high
T$_c$ superconductors 
may originate from a hidden long-range order, $d$-density
wave (DDW) \cite{nayak:ddw}.
This state is also related to the
staggered-flux state of Lee and Wen \cite{Lee-Wen:00},
but the latter is  dynamically fluctuating 
in their SU(2) gauge theory  of the cuprates.
According to Ref \cite{char1}, the pseudogap is a consequence of the
competition between two independent orders DDW and $d$-wave
superconductivity (DSC), which are transformable to each other 
in a 3 dimensional order parameter space and may coexist in the
underdoped cuprates.  A theory of DDW seems natural to account for a
possible quantum critical point near the optimal doping level 
that marks the onset of the pseudogap,  
put forward by  Tallon {\it et al} \cite{tallon} by examining the data of
photoemission, thermodynamic and transport properties, etc.  

This DDW scenario has recently attracted much attention about 
its nature and experimental
consequences \cite{wang,zhu,char2,Phillips,Honerkamp}.
Some investigations \cite{char2}
seem to indicate that various experiments 
in YBCO systems \cite{mook,sonier,miller} support this idea.  
The neutron scattering experiment \cite{mook} 
shows that the elastic signal around the in-plane
wave vector $Q=(\pi, \pi)$ in the underdoped YBCO appears well above
T$_c$.  
The $\mu SR$ experiment \cite{sonier} also confirms that
a small internal magnetic field appears above T$_c$ in the underdoped YBCO
but below T$_c$ in the
optimally doped samples.  Since internal magnetic fields are very weak and
spin fluctuations are too fast to couple with muon's spins, it is
reasonable to attribute them to DDW.  
However, both experiments also show that the magnetic 
signal is enhanced
when the temperature drops across T$_c$.
Such a behavior is not expected intuitively from the DDW picture,
since the DSC and DDW orders compete each other.
It thus
becomes quite interesting to understand how this temperature-dependent
puzzling behavior happens and, in
particular,
to see whether  it can be understood in terms of the existing 
self-consistent DDW mean-field theory
\cite{wang,zhu,char2,Phillips,Honerkamp}. 

In this article, we shall study the temperature dependence of the DDW
order at the mean-field level.  We find that it cannot give the
expected temperature dependence in the above experiments.  Instead,
our results show that the DDW order parameter is always suppressed
when the temperature drops below $T_c$, and that this is a robust
behavior, independent of the choice of parameters.  
Thermodynamic quantities (e.g. chemical potential, entropy, specific heat)
are also investigated. Their distinct temperature dependences are
discussed.

Consider the following mean-field Hamiltonian
\cite{wang,zhu,char2,Phillips,Honerkamp} 
\begin{eqnarray}
H_{MF}&=&\sum_{\avg{ij}\sigma} (-t_{\rm{eff}}-V_1 \chi_{ij}^*) 
c^\dagger_{i\sigma} c_{j\sigma}+h.c.\nonumber \\
&-& V_2\sum_{\avg{ij}\sigma}  \Delta_{ij} 
(c^\dagger_{i\uparrow}c^\dagger_{j\downarrow}
-c^\dagger_{i\downarrow}c^\dagger_{j\uparrow}
)+h.c. \label{eq:Hddw} \nonumber \\ 
&-&\mu\sum_{i\sigma} c^\dagger_{i\sigma} c_{i\sigma}+V_1
\sum_{\avg{ij}} 
\chi_{ij}^*\chi_{ij} +V_2 \sum_{\avg{ij}} \Delta_{ij}^*\Delta_{ij},
\nonumber
\end{eqnarray}
where $\avg{ij}$ indicates summation over the  nearest neighbors only. 
$\Delta_{ij}$ and the imaginary part of $\chi_{ij}$
play the role of
the DSC and DDW  order parameters, respectively. They are related
to the electron operators  via
$
\Delta_{ij} = \avg{c^\dagger_{i\uparrow}c^\dagger_{j\downarrow}
-c^\dagger_{i\downarrow}c^\dagger_{j\uparrow}}$ and
$\chi_{ij} = \avg{c^\dag_{i\sigma}c_{j\sigma}}$. 
$V_1$ and $V_2$ are positive in order to  have a nonzero DDW and
DSC order.
We have used an effective hopping amplitude $t_{\rm eff}=t \delta$
with $\delta$ the doping
concentration and $t$ the bare  hopping amplitude to take
account of the reduction of $t$ near
half-filling due to the strong Coulomb repulsion. Loosely speaking, 
the effective theory described above is equivalent to the fermion part
of the slave-boson mean-field theory of the
 $t$-$J$ model \cite{Kotliar,ubbens} 
in which the $J$-term is
decoupled into particle-hole and particle-particle channels with
different weights.

The  $t$-$J$ model at half filling has a (hidden) local
SU(2) symmetry \cite{Affleck++Fradkin:88}, which rotates
$(\Im \chi, \Re \Delta, \Im\Delta)$ as a {\bf 3}-vector. 
Thus the DDW($\pi$-flux) phase is degenerate with the DSC phase. 
Finite doping breaks this local SU(2) symmetry explicitly and favors
DSC order \cite{Kotliar}, because the Fermi surface nesting is destroyed.
Ubbens and Lee \cite{ubbens} showed that at
finite temperatures the DDW (flux)
state is  stable only  very  close to 
half-filling. The boundary between the DDW(flux) and DSC phases is of
the first order in nature.  Thus there is no coexistence phase.  In
the model we are considering, $V_1>V_2$ is needed to have DSC and DDW
coexist as pointed out in Ref.\cite{zhu}.  Although the
Heisenberg term  equally
favors DDW and DSC orders, the repulsion between the nearest
sites favors DDW over DSC. Hence the mean-field Hamiltonian above is
reasonably postulated.

\begin{figure} 
\centering\epsfig{file=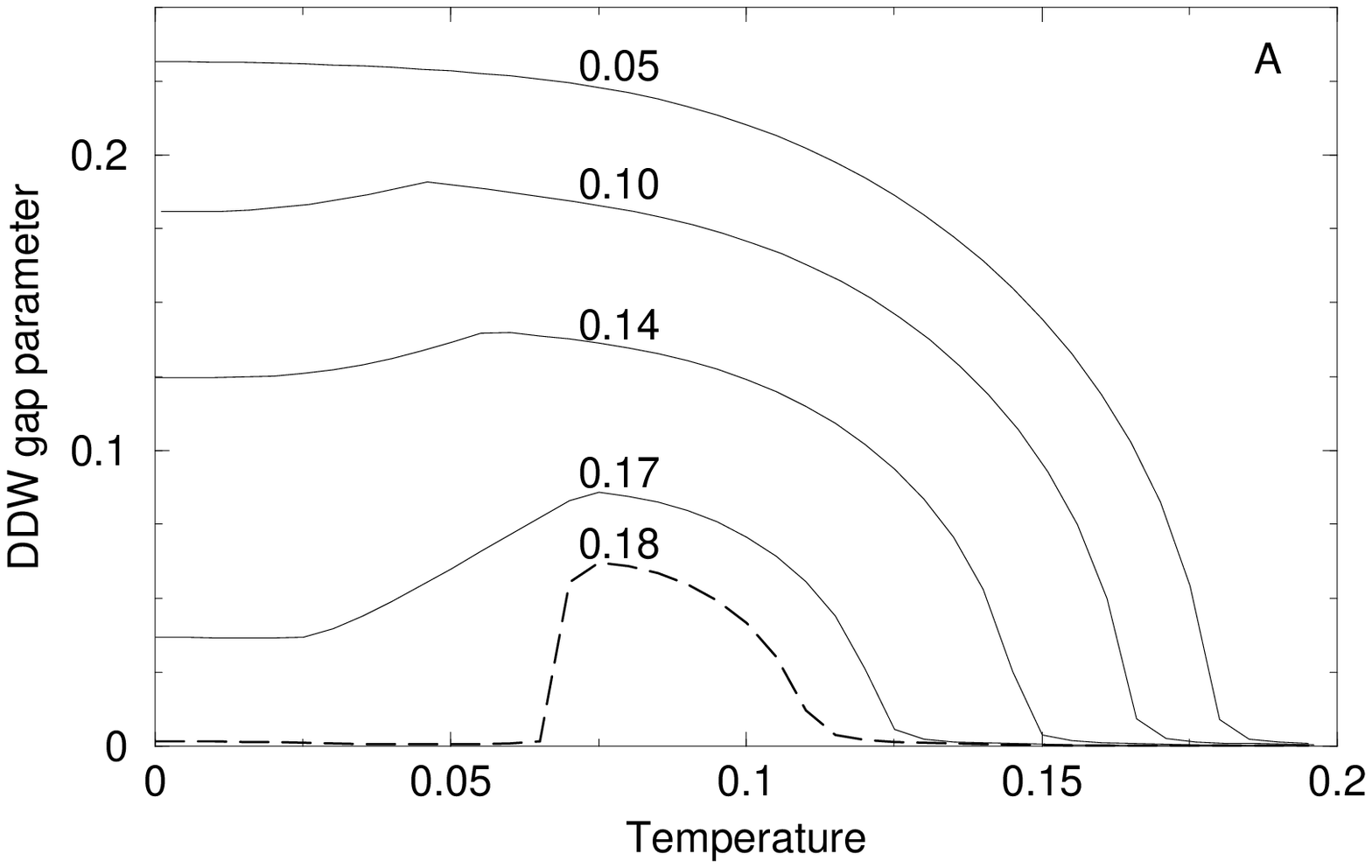,clip=1,width=\linewidth,angle=0}
\centering\epsfig{file=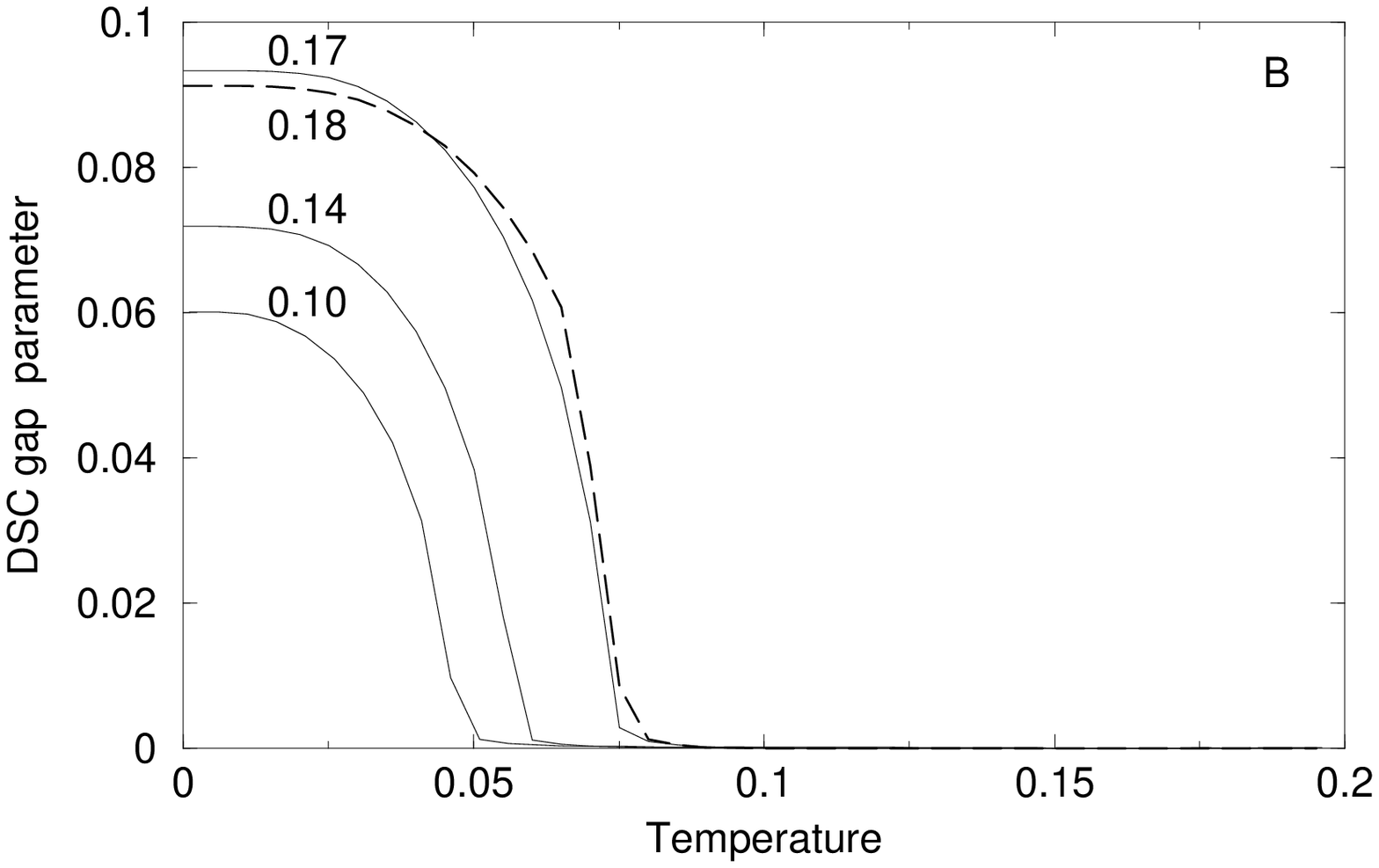,clip=1,width=\linewidth,angle=0}
\caption{(A) $W_{\rm DDW}$  vs temperature 
for various doping levels.
From top to bottom, $\delta=0.05,0.10,0.14,0.17,0.18$.
(B) $\Delta_{\rm DSC}$ vs temperature for various doping levels.
From bottom to top,
$\delta=0.10,0.14,0.18,0.17$. 
Curves of  $\delta=0.18$ are plotted with dashed lines
for the aid of eyes.
}\label{ordc}
\end{figure} 

Because the DDW order breaks the translational symmetry,
the Brillouin zone is reduced into one half and
the operators $(c_{k\uparrow}, c_{k+Q\uparrow},
c^\dagger_{-k\downarrow}, c^\dagger_{-k-Q\downarrow})$ are mixed
to give two branches of Bogoliubov quasiparticle excitations:
$
E(k)_{\pm} = 
\{ (-\mu \pm W_k)^2 +(2 V_2 \Delta \phi_k)^2 \}^{1/2}$,
where 
$
W_k  = \sqrt{ \epsilon_k^2+(2 V_1 \Im \chi \phi_k)^2}\,,
\epsilon_k =-(t_{\rm eff}+V_1 \Re \chi) \gamma_k$
and $ \phi_k = \cos k_x-\cos k_y, \gamma_k = \cos k_x+\cos k_y$
($\epsilon_k$ is the tight bond band energy.).
The corresponding self-consistent equations of $\Re\chi,\Im\chi$ and
$\Delta$ are:
\begin{eqnarray}
\Re \chi&=&  {1\over 2N} {\sum_k}^\prime
{-\epsilon_k \gamma_k \over W_k}
\Big\{ \tanh \left({\beta E_{k+} \over 2}\right) {-\mu +W_k\over E_{k+} }\nonumber\\
&& -\tanh \left({\beta E_{k-} \over 2}\right) {-\mu-W_k \over E_{k-} } \Big\},\nonumber
\\
\Im \chi&=& {1\over 2N}  {\sum_k}^\prime
{2V_1 \Im\chi ~  \phi_k^2 ~  \over W_k }
\Big\{ \tanh \left({\beta E_{k+} \over 2}\right) {-\mu +W_k\over E_{k+} }\nonumber\\
&&-\tanh \left({\beta E_{k-} \over 2}\right) {-\mu-W_k \over E_{k-} } \Big\},
\nonumber \\
 \Delta&=&{1\over 2N}  {\sum_k}^\prime
\Big\{ \tanh\left({\beta E_{k+} \over 2}\right)/ E_{k+}  
+\tanh \left({\beta E_{k-} \over 2}\right) \nonumber \\
&&/ E_{k-} \Big\} \times  2 V_2 \Delta \phi_k^2, \nonumber \\
\delta&=& {1\over N} {\sum_k}^\prime \Big\{ \tanh\left({\beta E_{k+} \over 2}
\right)
{-\mu+W_k\over E_{k+} }
+\tanh \left({\beta E_{k-} \over 2}\right) \nonumber \\
&& \times {-\mu-W_k\over E_{k-}} 
\Big\}, \nonumber
\end{eqnarray}
where the summation is restricted within the reduced Brillouin zone
and $\beta=1/T$.
Below we denote the energy gaps $\Delta_{\rm DSC}= 2 V_2 \Delta$ and
$W_{\rm DDW}= 2 V_1 \Im \chi$ for the DSC and DDW orders, respectively.

After solving the self-consistent equations at 
$V_1=0.38$ and  $V_2=0.25$ with energy scale set as
$t\equiv 1$, we obtain the dependence of the $\Delta_{\rm DSC}$
and $W_{\rm DDW}$ gap vs  doping $\delta$ at zero temperature 
as shown in Fig.~1 of Ref.~\cite{char1}.
$\Delta_{\rm DSC}$  begins to develop after $\delta>0.06$ and
reaches maximum at $\delta \approx 0.18$.
$W_{\rm DDW}$ also drops to zero around there.
The phase diagram of temperature  vs $\delta$ is similar to Ref \cite{zhu},
and we shall not reproduce it here. 

There exists a coexisting region of both orders.
However,  in this region the behavior
$W_{\rm DDW}$ vs temperature ($T$) is subtle (see
Fig.~\ref{ordc}A).  In comparison, Fig.~\ref{ordc}B  shows how 
$\Delta_{\rm DSC}$ varies with $T$.
For very low doping ($\delta$=0.05) where $\Delta_{\rm DSC}=0$, 
$W_{\rm DDW}$ is monotonically enhanced when $T$ is reduced.
When the coexisting region is entered, 
$W_{\rm DDW}$ becomes maximum  around 
T$_c$ where $\Delta_{\rm DSC}$ starts developing.
This feature is general for competing  orders,
which also happens in  the competition of $s$ and
$d$-wave pairing orders \cite{sd}.
When either order develops, it generates a gap  
near the Fermi surface. Consequently, it becomes difficult for the other
to form.
When $T$
slightly drops from T$_c$, $\Delta_{\rm DSC}$ increases fast as  
$ (1-T/T_c)^{1/2}$. 
$W_{\rm DDW}$  loses more weight to DSC than it 
gains from lowering temperature.
When $T$ drops well below
 $T_c$,  $\Delta_{\rm DSC}$  increases very slowly and 
$W_{\rm DDW}$   changes little as well.
In the underdoped  region, 
$W_{\rm DDW}$  still has a substantial residual value at 
$T=0$~K, which gets significantly
reduced and may even become zero 
near the optimal doping.
When $\delta=0.18$ in Fig.~\ref{ordc}A, $W_{\rm DDW}$
only exists in a small range around  T$_c$.
In other words, something like the re-entry phenomenon occurs
here  with varying temperatures: 
$W_{\rm DDW}$ vanishes at $T=0$K, 
begins to increase from a finite temperature  to $T_c$, and 
then decreases to zero again when $T>T_c$.
We never see $W_{\rm DDW}$  developing with decreasing
$T$ within the superconducting region.
These are just opposite to what the neutron scattering
and $\mu$SR experiments  indicated.

We have to take it seriously why experimental signals are enhanced when
$T<T_c$.  There seem to be only two possibilities.  The first one is that
these signals are really related to the DDW order. Then a
modified  mean-field theory is needed for a
mechanism that the two orders can somehow enhance each other around
$T_c$.  Or these signals have other origins such as
from spin, then they cannot be used as evidence for the existence of
the DDW order.

\begin{figure} 
\centering\epsfig{file=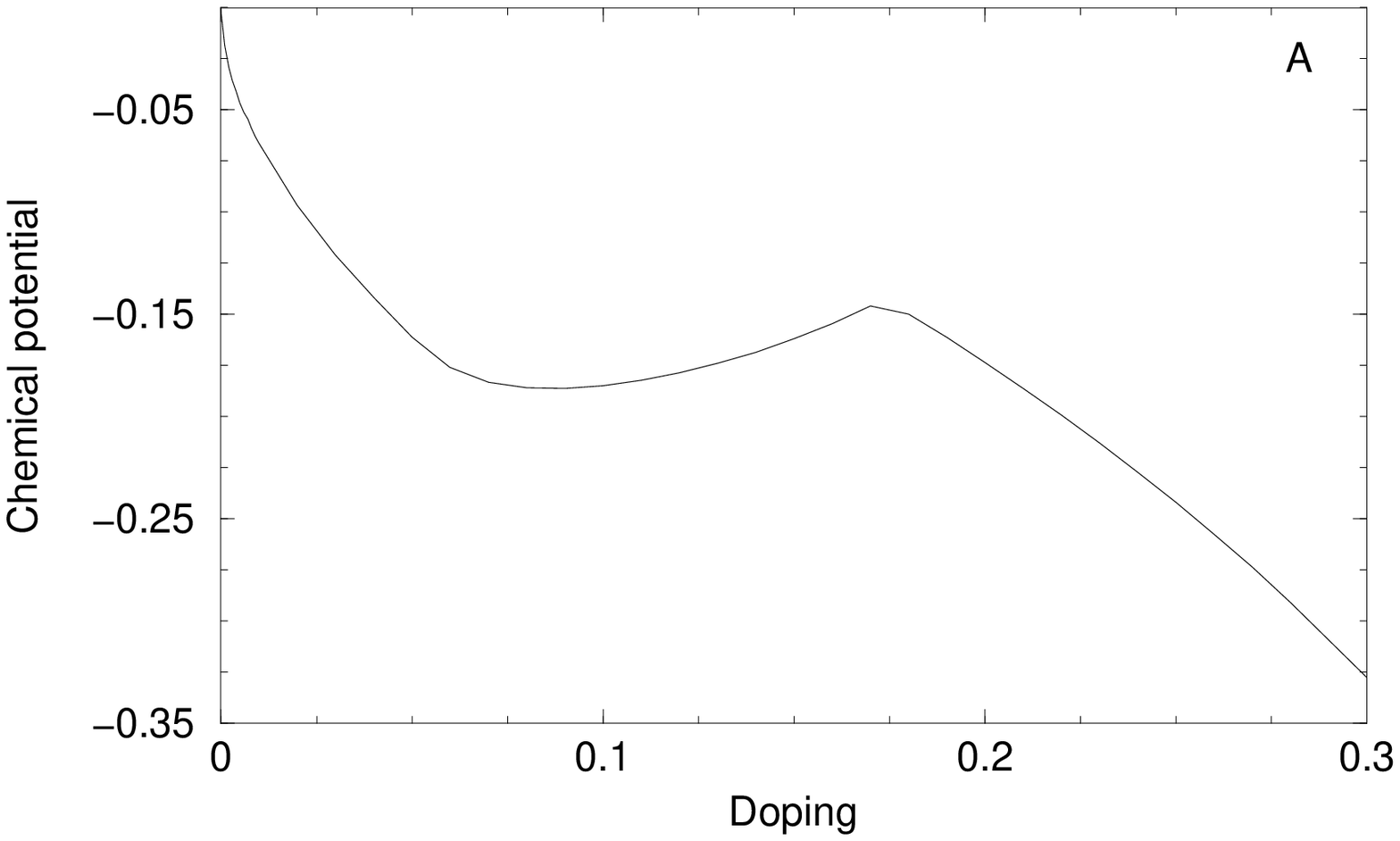,clip=1,width=\linewidth,angle=0}
\centering\epsfig{file=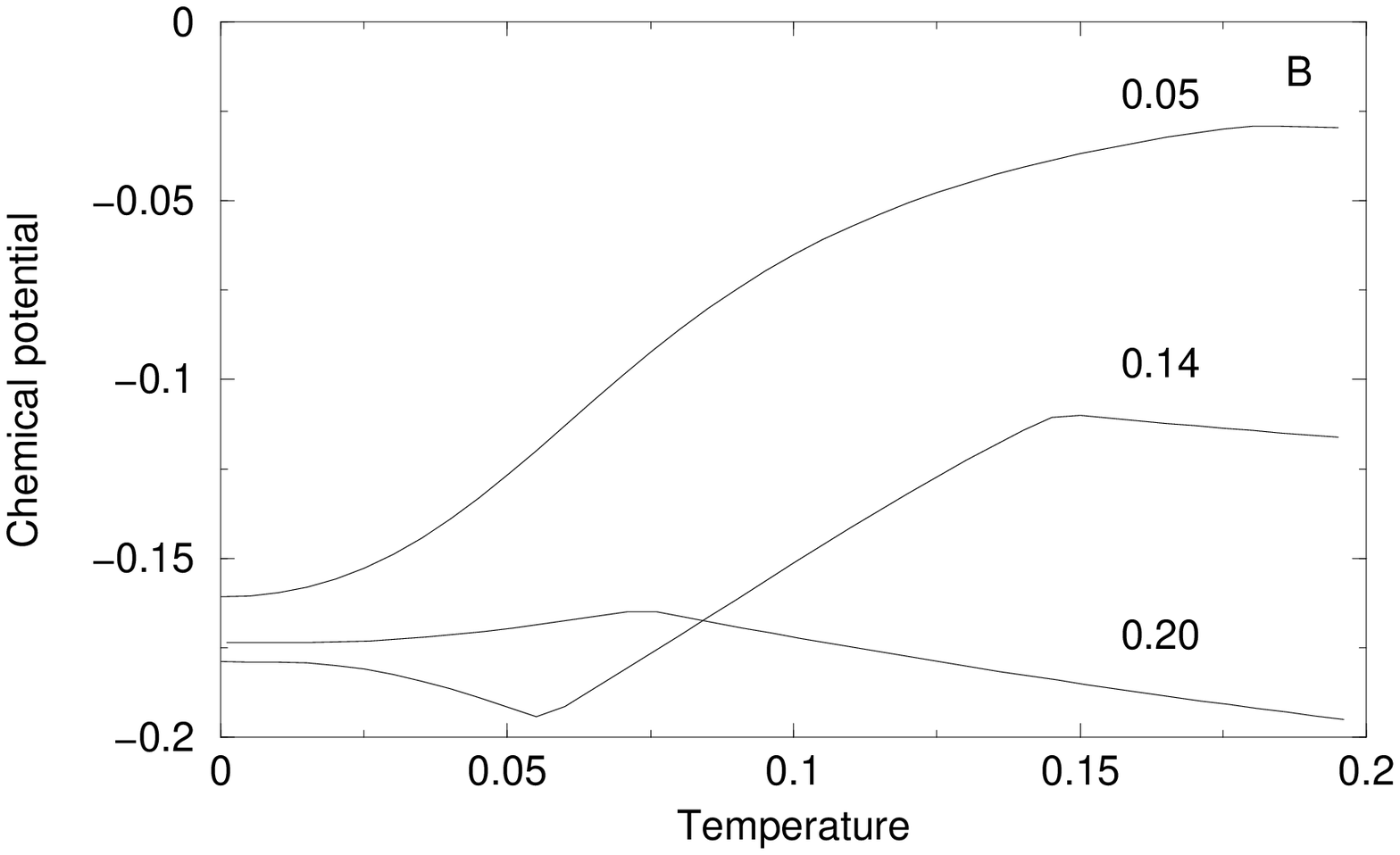,clip=1,width=\linewidth,angle=0}
\caption{A. Chemical  potential vs doping at zero temperature.
B. Chemical potential vs temperatures at characteristic doping levels.
$\delta= 0.05,  0.14, 0.20$.
}\label{chem}
\end{figure}

The behavior of chemical potential $\mu$ is also interesting.
In Fig.~\ref{chem}A,  we show the dependence of $\mu$ with  $\delta$
at zero temperature.
$\mu$ decreases with increasing $\delta$  when $\delta \lesssim 0.06$,  
increases slowly in the underdoped region
and drops quickly in the overdoped region.
This can be understood as follows. The energy curve of
quasiparticles is cone-shaped in the momentum space with a Fermi pocket
near $(\pi/2,\pi/2)$. After the onset of  $\Delta_{\rm DSC}$,
$W_{\rm DDW}$ drops faster with increasing $\delta$, 
so the density of states (DOS) increases.
This keeps $\mu$ roughly unchanged while increasing doping.
In the underdoped region $\partial \mu / \partial n$
is small and becomes even negative, 
which means that the 
charge instability may develop here.
Photoemission experiments \cite{ino} show that
$\mu$ is almost fixed 
at the undoped value upon increasing $\delta$ 
in the underdoped region.
Our result agrees with that qualitatively, 
but $\mu$ is not
fixed at the value of the undoped case.

Fig.~\ref{chem}B shows the temperature dependence of $\mu$ 
at three doping levels: low (nonsuperconducting) doping 
($\delta$=0.05), underdoped($\delta$=0.14), and overdoped($\delta$=0.20).
In the low doping region,
$\mu$ increases with temperature increasing.
$W_{\rm DDW}$ is weakened by temperature while
$\mu$ increases to fix the particle number.
In the underdoped region,
$\mu$'s behavior is subtle.
It first drops when $T<T_c$, then increases when $T_c<T<T_{\rm DDW}$, 
and drops again when $T>T_{\rm DDW}$, where $T_{\rm DDW}$ is the onset
temperature for DDW.
This can be explained by the temperature dependence of $W_{\rm DDW}$ 
(Fig.~\ref{ordc}).
$W_{\rm DDW}$ is enhanced (or weakened) by increasing temperature
within the range $T<T_c$ (or $T_c<T<T_{\rm DDW}$).
Thus $\mu$ first drops  and then increases.
After $T$ passes over $T_{\rm DDW}$, 
$\mu$ drops as an ordinary Fermi gas  behaves.
In the overdoped region, $W_{\rm DDW}=0$.
Thus $\mu$ is almost fixed when $T<T_c$ but drops when $T>T_c$.

\begin{figure} 
\centering\epsfig{file=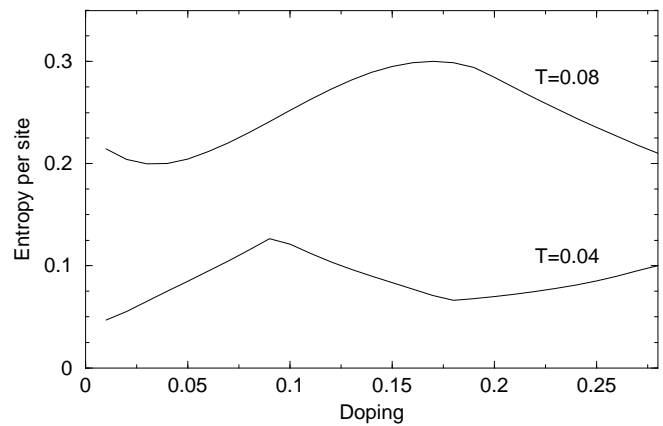,clip=1,width=\linewidth,angle=0}
\caption{Entropy per lattice site vs doping at fixed temperatures.
From top to bottom, T=0.08, 0.04.
}\label{entr}   
\end{figure}

The DDW order also has important effects on the entropy per site $S$
vs $\delta$, as  shown  in Fig \ref{entr}.
The first curve is at the high temperature 
where  $\Delta_{\rm DSC}=0$ and only $W_{\rm DDW}$ exists.
In the underdoped region, $S$ decreases when $\delta$ is reduced,
because $W_{\rm DDW}$  reduces the low energy DOS. 
In the overdoped region, $S$ drops when $\delta$ increases, 
which is the standard Fermi liquid behavior.
Thus $S$ reaches maximum near the  optimal doping.
This agrees with  experimental results of Loram {\it et al} 
\cite{loram}.
Very close to half filling, the hopping amplitude is reduced.
As a result, the band width is reduced and 
DOS  is enhanced.
This effect tends to increase $S$.
Simultaneously, the DDW order is enhanced by lowering doping,
which has an effect to decrease $S$.
At high temperatures where $W_{\rm DDW}$ is small,
the first effect may overcome the second one
and thus $S$ increases when $\delta$  decreases.
This phenomenon  is absent at the low temperature
where $W_{\rm DDW}$ is large, which is also shown in Ref \cite{loram}.
Let us increase $\delta$ at the lower temperature as in the 
bottom curve.
$S$ increases  at very low doping levels,
since  only $W_{\rm DDW}$ exists. 
In the coexistence or underdoped region,
$S$ drops because  $\Delta_{\rm DSC}$   develops.
After passing the optimal point,  
$\Delta_{\rm DSC}$   decreases and thus makes $S$ increase again.

There is also specific heat anomaly 
at the onset of $W_{\rm DDW}$, as shown in Fig 5.
The jump of the specific heat coefficient $\gamma(T)=C(T)/T$ 
is large and can be compared with those 
at the onset of $\Delta_{\rm DSC}$.
The jump  at the DDW transition is larger 
at the more underdoped side,
while that at the DSC transition's behavior is just opposite.
However, the former is not seen in experiments
and is a difficulty for the DDW scenario.
It was  argued that disorder
removes the sharp  transition and turns it into a smooth crossover
in Ref. \cite{char1}.
Ref. \cite{kee} suggests that a negatively large $\mu$
can weaken the jump by destroying the nesting of Fermi surface.
However, from the Fig.~2B,  $\mu$ increases rapidly when
T$\rightarrow$ T$_{DDW}$,
and  $|\mu|$ is much smaller at $T_{DDW}$ than 
at zero temperature, especially at the low doping region.
It is still difficult to understand why   $\mu$ can be  negatively large.

\begin{figure} 
\centering\epsfig{file=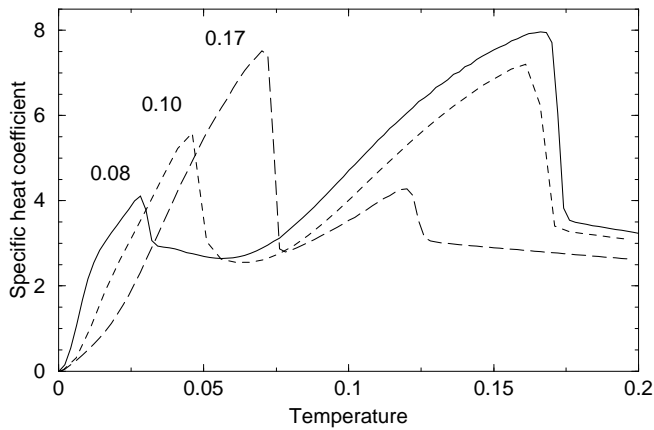,clip=1,width=\linewidth,angle=0}
\caption{ Specific heat coefficient $\gamma(T)$ vs $\delta$
in the  underdoped and optimal region,
from left to right $\delta=$0.08, 0.10, 0.17.
}\label{ord}
\end{figure}

At last, we briefly discuss  the condensation energy U$_0$. In the case
of the pure DSC state, electrons near $(\pi,0)$ contribute much to   U$_0$
and those near $(\pi/2,\pi/2)$ contribute little.
However, in the DDW scenario,
there is a pre-existing $W_{\rm DDW}$ in the pseudogap region by assumption.
Upon doping, the Fermi surface is a small pocket, 
which has not been seen in experiments yet.
The same $d_{x^2-y^2}$ symmetry makes the vicinity of 
$(\pi,0)$ already far
below the  Fermi surface. 
Then  paring can not affect them as significantly as in the case of pure DSC,
thus  U$_0$ is reduced.
Let us study the contribution 
to  U$_0$  along the curve of the minimum gap 
form the direction $(\pi/2, \pi/2)\rightarrow (\pi,0)$.
At beginning, it is proportional to $\Delta_{\rm DSC} \phi_k$, where
$\phi_k= \cos k_x-\cos k_y$.
After passing the point $k_0$ where 
the end of the Fermi pocket lies, 
the dependence  changes into
$\phi_k(\sqrt{W_{\rm DDW}^2 (1-\phi_{k0}/\phi_k)^2+\Delta_{\rm DSC}^2}-
W_{\rm DDW} (1-\phi_{k0}/\phi_k) )$.
If $W_{\rm DDW}$ is large, the slope becomes softer and  
an apparent kink develops.
This kink can be testified by studying the retreat of the leading edges
of the high resolution ARPEPS data
deep in the superconducting region relative to those in the pseudogap region.
If the pseudogap is caused by pair fluctuations, such kink will not exist. 

In summary, we studied the DDW and DSC order parameters' dependence
with temperature and  thermodynamic quantities in detail by solving 
the mean-field Hamiltonian self-consistently.
The DDW order is suppressed when temperature drops
below T$_c$ in the underdoped region because of their competing nature.
The disagreement with experimental results was discussed.
Behaviors of the chemical potential, entropy and specific
heat  with temperature and doping are investigated. 
The increase of chemical potential is predicted
when the temperature increases in the pseudogap region.
We also showed the decrease of entropy when doping decreases
in the underdoped region.
The distribution of  condensation energy in the momentum space
has a kink along the direction form $(\pi/2, \pi/2)$ to $(\pi,0)$.
These features may be used in experiments to testify the DDW scenario.

We thank E. Fradkin, A. J. Leggett, P. Phillips, 
J. L. Tallon and J. W. Loram for their kind help.
This work is supported  by NSF grant
DMR98-17941 and DMR01-32990 at UIUC.
 W.V.L. is also supported  in part by funds
provided by the U.S. Department of Energy (DOE) under cooperative
research agreement \#DF-FC02-94ER40818 at MIT.


\begin{thebibliography}{10}

\bibitem{history} B. I. Halperin {\it et al}, Solid State
        Phys. {\bf 21}, 116 (1968);
H. J. Schulz, {\it ibid} {\bf 39}, 2940 (1989);
E. Cappelluti {\it et al}, {\it ibid} {\bf 59}, 6475(1999);
L. Benfatto {\it et al}, Eur. Phys. J. B {\bf 17}, 95(2000);
B. Dora {\it et al}, {\it ibid} {\bf 22}, 167(2001).


\bibitem{char1}
S. Chakravarty et al.,
Phys. Rev. B {\bf 63}, 094503-1 (2001). 

\bibitem{nayak:ddw} C. Nayak and F. Wilczek, cond-mat/9510132 (unpublished);
        C. Nayak, {\it Phys. Rev. B} {\bf 62}, 4880 (2000).


\bibitem{Lee-Wen:00} P.A. Lee and X. G. Wen, Phys. Rev. B {\bf 63}, 224517
(2001).

\bibitem{tallon}  
J. L. Tallon and J. W. Loram, Physica C {\bf 349}, 53 (2001).

\bibitem{wang}
Q. H. Wang {\it et al}, Phys. Rev. Lett.{\bf 87}, 077004 (2001).


\bibitem{zhu} J. X. Zhu {\it et al} ,
Phys. Rev. Lett {\bf 87}, 197001 (2001).

\bibitem{char2} 
S. Chakravarty {\it et al}, Int. J. Mod. Phys. B {\bf 15}, 2901 (2001);
S. Tewari {\it et al}, Phys. Rev. B {\bf 64}, 224516 (2001).

\bibitem{Phillips} 
T. Stanescu {\it et al}, Phys. Rev. B {\bf 64}, 220509 (2001).

\bibitem{Honerkamp} C. Honerkamp {\it et al},
J. Phys. Cond. Matt. {\bf 13}, 11669 (2001).

\bibitem{mook}
H. A. Mook {\it et al}, Phys. Rev. B {\bf 64}, 012502-1(2001).

\bibitem{sonier}
J. E. Sonier  et al,
Science {\bf 292}, 1692 (2001).  

\bibitem{miller}
R. I. Miller et al,
cond-mat/0111550

 
\bibitem{Kotliar}
G. Kotliar and J. Liu, Phys. Rev. B {\bf 38}, 5142(1988).

\bibitem{ubbens} M. U. Ubbens and P. A. Lee, Phys. Rev. B 46, 8434 (1992).


\bibitem{Affleck++Fradkin:88} I. Affleck, et al., Phys. Rev. B {\bf
38}, 745 (1988); E.  Dagotto, E. Fradkin and A. Moreo, Phys. Rev. B
{\bf 38}, 2926 (1988).

\bibitem{sd} J. X. Zhu et al., Phys. Rev. B {\bf 57},
13410 (1998).

\bibitem{ino} A. Ino et al., Phys. Rev. Lett {\bf 79}, 2101 (1997).

\bibitem{loram}
J. W. Loram {\it et al},
J. Phys. Chem. Solids {\bf 59}, 2091(1998);
{\it  ibid} {\bf  62}, 59(2000).

\bibitem{kee} H. Y. Kee, Y. B. Kim, cond-mat/0111461.
\end{thebibliography}
\end{document}